# Parallelism Via Concurrency at Multiple Levels Computer Architecture

**Kamran Latif,**
**University of Lahore**
*Email: kamranlatif@gmail.com*

### Abstract

*In this paper we examine the key elements determining the best performance of computing by increasing the frequency of a single chip, and to get the minimum latency during execution of the programs to achieve best possible output. It's not enough to provide concurrent improvements in the HDW as Software also have to introduce concurrency in order to exploit the parallelism. The software parallelism is defined by the control and data dependency of programs whereas HDW refers to the type of parallelism defined by the machine architecture and hardware multiplicity.*

***Key Words***: Hardware (HWD), Instruction Level Parallelism (ILP), Graphical user interface (GUI), chip multiprocessor (CMP), Digital signal processor (DSP), Thread level Parallelism (TLP),

### 1. Introduction

By implementing the Moore's law stating that :( *The number of transistors on a chip wills roughly double every 18 months):* the electrical engineers empowered the computer design engineers to create the complex microprocessors. To improve the performance in past years many efforts were made to enhance the HDW which are slightly discussed in paper? As now it seems that the computer industry is already at the advance level and the design engineers are getting short of the concepts regarding the hardware improvement so the reforms are also supposed to be made in the software to compensate the improvement process.

### 2. Frequency Scaling

In computer architecture, frequency scaling (also known as frequency ramping) so as to achieve performance gains. Frequency ramping was the dominant force in commodity processor performance increases from the mid-1980s until roughly the end of 2004. The effect of processor frequency on computer speed can be seen by looking at the equation for computer program runtime

$$\text{Runtime} = \frac{Instructions}{Program} X \frac{Cycles}{Instruction} X \frac{Time}{Cycle}$$

(Where instructions per program are the total instructions being executed in a given program, cycles per instruction are a program-dependent, architecture-dependent average value, and seconds per cycles is by definition the inverse of frequency. *An increase in frequency thus decreases runtime).*

### 3. Automatic Exploitation Of ILP

An instruction stream has a combination of memory and computational operations. When the memory operations are issued, there can be some independent computational instructions that can be executed. But additional hardware is required to find these independent instructions and schedule them so that no data hazard occurs. This technique is called automatic exploitation of Instruction Level Parallelism (ILP). Computer designers have tried hard to improve performance of the programs through implicit ILP without changing anything in the software. Because of implicit ILP, the programmers were not required to have a detailed knowledge of the hardware. The programmer used to wait until a more powerful machine is available in the market and then the software mysteriously got faster.

However, the industry is reaching the limits of increasing the frequency of processors to execute more instructions in one cycle.

### 4. Concurrency in Hardware.

Concurrency is a property of systems in which several computations are executing simultaneously, and potentially interacting with each other. The computations may be executing on multiple cores in the same chip, preemptively time-shared threads on the same processor, or executed on physically separated processors or even hundreds of cores on a single chip. We can do multiple things at once (parallelism, multitasking). But concurrency introduces non-determinism: the exact order in which things are done (the schedule) is not known in advance. It is often the case that some schedules will lead to correct outcomes and vice versa. Today concurrency has leaked into application programs. Rich GUIs must respond to user events while other activities are in progress. In short parallelism is not possible unless concurrency is introduced explicitly in programs.

## 5. Concurrency In Software

The concurrency is mainly a software transformation and the problem is not building multicore hardware but programming it in a way that lets main stream applications benefit from the continued exponential growth in CPU performance. The software industry needs to get back into the state where existing applications run faster on new hardware.

## 6. Multi-core Systems

A multi-core processor is a single computing component with two or more independent actual central processing units (called "cores"), which are the units that read and execute program instructions. The instructions are ordinary CPU instructions such as add, move data, and branch, *but the multiple cores can run multiple instructions at the same time*, increasing overall speed for programs amenable to parallel computing. Manufacturers typically integrate the cores onto a single integrated circuit die (known as a chip multiprocessor or CMP), or onto multiple dies in a single chip package.

Multi-core processors are widely used across many application domains including general-purpose, embedded, network, digital signal processing(DSP), and graphics.

## 7. Parallel Computing

Parallel computing is a form of computation in which many calculations are carried out simultaneously, operating on the principle that large problems can often be divided into smaller ones. Which are then solved concurrently ("in parallel"). There are several different forms of parallel computing: bit-level, instruction level, data, and task parallelism. Parallelism has been employed for many years, mainly in high-performance computing.

### 7.1 Amdahl's law 1960

Amdahl's law, also known as Amdahl's argument is used to find the maximum expected improvement to an overall system when only part of the system is improved. It is often used in parallel computing to predict the theoretical maximum speed up using multiple processors.

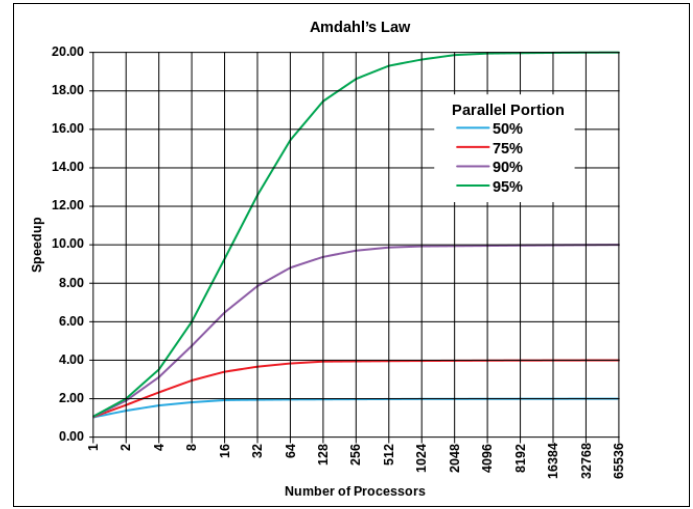

Fig. 1. Example Using multi-processers. *(parallel Computing)*

*The speedup of a program using multiple processors in parallel computing is limited by the sequential fraction of the program. For example, if 95% of the program can be parallelized, the theoretical maximum speed up using parallel computing would be 20 × as shown in the diagram, no matter how many processors are used.*

#### 7.1.1 Speed up in Sequential Program

The maximum speed up in an improved sequential program, where some part was speed up $p$ and times is limited by inequality.

Maximum Speedup $\leq \frac{P}{1+f \times (p-1)}$

Where $f$ $(0<f<1)$ is the fraction of time (before the improvement) spent in the part that was not improved.

- If part B is made five times faster ($p = 3$), $t_A = 3$, $t_B = 1$, and $f = t_A / (t_A + t_B) = 0.75$ then
  
  Maximum Speedup $\leq \frac{5}{1+0.75 \times (5-1)} = 0.71$

- If part A is made to run twice as fast $p = 2$ $t_B = 1$, $t_A = 3$ and $f = t_B /(t_A + t_B) = 0.25$
  
  Then
  
  Maximum Speedup $\leq \frac{3}{1+0.25 \times (2-1)} = 2.4$

Therefore, making Part A twice as fast is better than making part B five times faster. The percentage improvement in speed can be calculated as

Percentage improvement = $(1 - \frac{1}{\text{Speedup factor}}) \times 100$

Improving part A by a factor of two will increase overall program speed by a factor of 2.4 which makes it 58.33% faster than the original computation.

However, improving part B by a factor of five, which presumably requires more effort, will only achieve an overall speedup factor of 0.71, which makes it 40.84% faster.

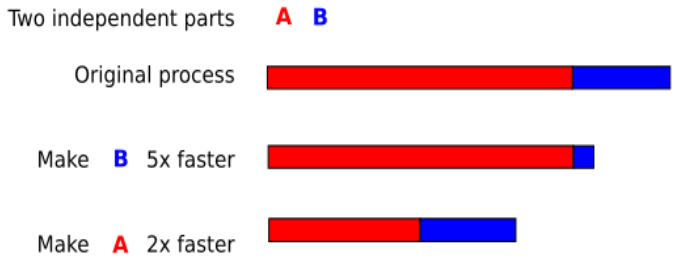

Fig 2 Example Speed up Sequential Program Amdahl's law

### 7.2 Gustafson's law

Gustafson's law is a law in computer science which says that computations involving arbitrarily large data sets can be efficiently parallelized. Gustafson's Law provides a counterpoint to Amdahl's law, which describes a limit on the speed-up that parallelization can provide. Given a fixed dataset size. Gustafson's law was first described

S (P) = P – α x (P -1)

Where P is the number of processors, S is the speedup, and α non-parallelizable fraction of any parallel process.

Gustafson's law addresses the shortcomings of Amdahl's law, which does not fully exploit the computing power that becomes available as the number of machines increases

Gustafson's Law proposed that programmers manage to set the size of problems to use the available resources to solve problems within a practical fixed time.

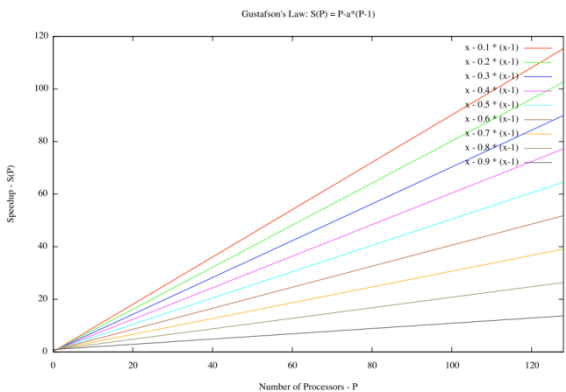

### Comparison

### Amdahl's law

*Suppose a car is traveling between two cities 60 miles apart, and has already spent one hour traveling half the distance at 30 mph. No matter how fast you drive the last half, it is impossible to achieve 90 mph average before reaching the second city. Since it has already taken you 1 hour and you only have a distance of 60 miles total; going infinitely fast you would only achieve 60 mph.*

### Gustafson's Law

*Suppose a car has already been traveling for some time at less than 90mph. Given enough time and distance to travel, the car's average speed can always eventually reach 90mph, no matter how long or how slowly it has already traveled. For example, if the car spent one hour at 30 mph, it could achieve this by driving at 120 mph for two additional hours, or at 150 mph for an hour, and so on.*

Fig 3 Example Speed up Gustafson's law

## 8. Thread Level Parallelism (TLP)

Thread Level Parallelism is the process of splitting a program into independent parts and having these parts run side by side as "threads". This can be having separate independent programs running together or having separate activities inside the same programs. There are two basic methodologies in Thread Level Parallelism, fine-*grained multithreading* and course-grained multithreading, TLP provides each thread with its own copy of the Program Counter (PC), register files, etc. Multithreading improves the throughput of computers that are running multiple programs and the execution time of multi-threaded programs is decreased.

## 9. Conclusion

Concurrency also opens the possibility of new, richer computer interfaces and far more robust functional software. This requires a new burst of imagination to find and exploit new uses for the exponentially increasing potential of new processors. To enable such applications, programming language designers, system builders, and programming tool creators need to start thinking seriously about parallelism and find techniques better than the low-level tools of threads and explicit synchronization that are today's basic building blocks of parallel constructs that more clearly express a programmer's intents , so that that architecture of a program is more visible, easily understood and verifiable by the tools.

### References

[1] Irfan Uddin Advances in Computer Architecture University of Amsterdam, the Netherland 2013

[2] Saxena, V ; IBM Res. - India, New Delhi, High Performance Computing (HiPC), 2010 International Conference paper at Dona Paula.

[3] Vikas Aggarwal, HPCC Random Access Benchmark for Next Generation Supercomputers.

[4] HERB SUTTER, Software and the Concurrency Revolution Sep 2005

[5] JAMES LARUS is a senior researcher at Microsoft Research associate professor at the University of Wisconsin-Madison,(2005)